\documentclass[twocolumn,showpacs,prl]{revtex4}

\usepackage{graphicx}
\usepackage{dcolumn}
\usepackage{bm}

\begin{document}

\title{Thermally Activated Processes in Polymer Glasses}

\author{V. Parihar, D. Drosdoff, A. Widom}
\affiliation{Physics Department, Northeastern University, Boston MA 02115, USA}
\author{Y.N. Srivastava}
\affiliation{Physics Department \& INFN, University of Perugia, Perugia, IT}

\begin{abstract}

A derivation is given for the Vogel-Fulcher-Tammann thermal activation law  
for the glassy state of a bulk polymer. Our microscopic considerations involve 
the entropy of closed polymer molecular chains (i.e. polymer closed strings). 
For thin film polymer glasses, one obtains open polymer strings in that the 
boundary surfaces serve as possible string endpoint locations. The 
Vogel-Fulcher-Tammann thermal activation law thereby holds true for a bulk 
polymer glass but is modified in the neighborhood of the boundaries of 
thin film polymers.  

\end{abstract}

\pacs{61.43.Fs, 61.20.Lc, 64.70.Pf, 71.55.Jv, 81.05.Kf}

\maketitle

\section{Introduction \label{intro}}

There has been considerable recent interest on the general dynamics of the glass 
transitions in bulk polymer 
systems\cite{Bendler:2005,Donth:2001,Casalini:2004,Debenedetti:1997,Debenedetti:2001,Tanaka:2003,Kirkpatrick:1989}. 
A central experimental law which controls the rate of transition was long ago 
formulated by Vogel, Fulcher and Tammann\cite{Vogel:1921}; The empirical VFT law 
of transition rates reads  
\begin{equation}
\Gamma = \nu \exp\left\{-\left[\frac{\Phi }{k_B(T-T_0)}\right] \right\},
\label{intro1}
\end{equation}
wherein \begin{math} \Phi  \end{math} is the free energy of thermal activation. 
The VFT thermal activation law is quite similar to the well known 
Arrhenius\cite{Arrhenius:1889} thermal activation law except for the temperature 
singularity in the denominator on the right hand side of Eq.(\ref{intro1}).
The singularity occurs at a dynamical temperature \begin{math} T_0 \end{math} which 
is somewhat lower than the thermodynamic glass transition temperature 
\begin{math} T_g \end{math}. The singularity is thereby never quite attained. 
Nevertheless, the critical slowing down of the VFT Eq.(\ref{intro1}) 
is experimentally well obeyed in bulk polymer glasses. There exist 
somewhat different physical views\cite{Saslow:1988,Garcia:1989,Bendler:2001} 
as to why the VFT law might theoretically be true. Nevertheless 
there is presently no agreed upon theory of Eq.(\ref{intro1}). 

Our purpose is to derive the VFT thermal law through the following quite simple quantum mechanical 
considerations. The transition rate per unit time for an activated process involves 
an absolute squared transition amplitude (matrix element) times a density of final states.
The logarithm of the density of final states represents the final entropy. Thus, 
the quantum mechanical rule for computing transition rates is that
\begin{equation}
\Gamma =\nu_\infty \exp[S(E)/k_B],
\label{intro2}
\end{equation}
wherein \begin{math} S(E) \end{math} is the entropy of activation to a state with energy 
\begin{math} E  \end{math}. The theoretical problem is to deduce the nature of the 
excitations\cite{Shenoy:2000,Adam:1965,Blackburn:1994} 
and compute the entropy of activation from the the logarithm of the final state 
phase space magnitude  
\begin{equation}
S(E)=k_B \ln \Omega (E).
\label{intro3}
\end{equation}

\begin{figure}[tp]
\scalebox {0.55}{\includegraphics{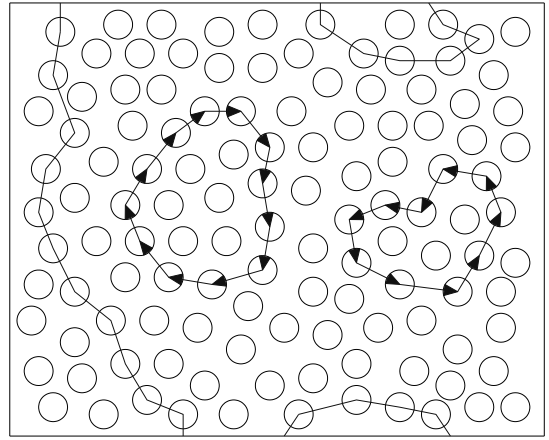}}
\caption{\it Schematically shown are two closed mobile chains of polymers which occur 
as ``bulk closed strings''. Also schematically shown are three polymer chains which 
begin and end on the surface boundaries of the bulk polymer and constitute ``open strings''.}
\label{fig1}
\end{figure}

The polymer glass excitation configurations\cite{Donati:1998,Blackburn:1996,arndt:1997} 
pictured in FIG.\ref{fig1} are of two 
types: (i) There are - in the bulk of the polymer - closed chains of atoms referred to as {\em closed} strings. (ii) Also, there are open polymer chains which begin and end 
on the boundary surfaces of the bulk polymer and are referred to as {\em open} 
strings. It will be shown below that the closed strings have an entropy obeying 
the VFT thermal activation Eq.(\ref{intro1}). On the other hand, the open string configurations 
with end points in the neighbourhood of surface boundaries obey shifted thermal activation laws. 

The distinction between the thermal activation properties of open and 
closed strings is crucial for an understanding
of surface effects which are of importance for thin films\cite{Masson:2002,Wu:1995,deGennes:2000}.
The VFT thermal activation 
law holds only for the bulk polymer. By contrast, the dynamical sinularity temperature
\begin{math} T_0 \end{math} decreases as the ratio of boundary surface 
are to the bulk volume,  \begin{math} L^{-1} \end{math}, increases. Consequently, the singularity 
temperature is sharply lowered\cite{Varnik:2002,Wallace:1995,Forrest:1997,Colby:2000} 
when \begin{math} L \end{math} is decreased to a few nanometers. 

\section{Entropy of Closed Mobile Chains\label{closed}}

\begin{figure}[tp]
\scalebox {0.7}{\includegraphics{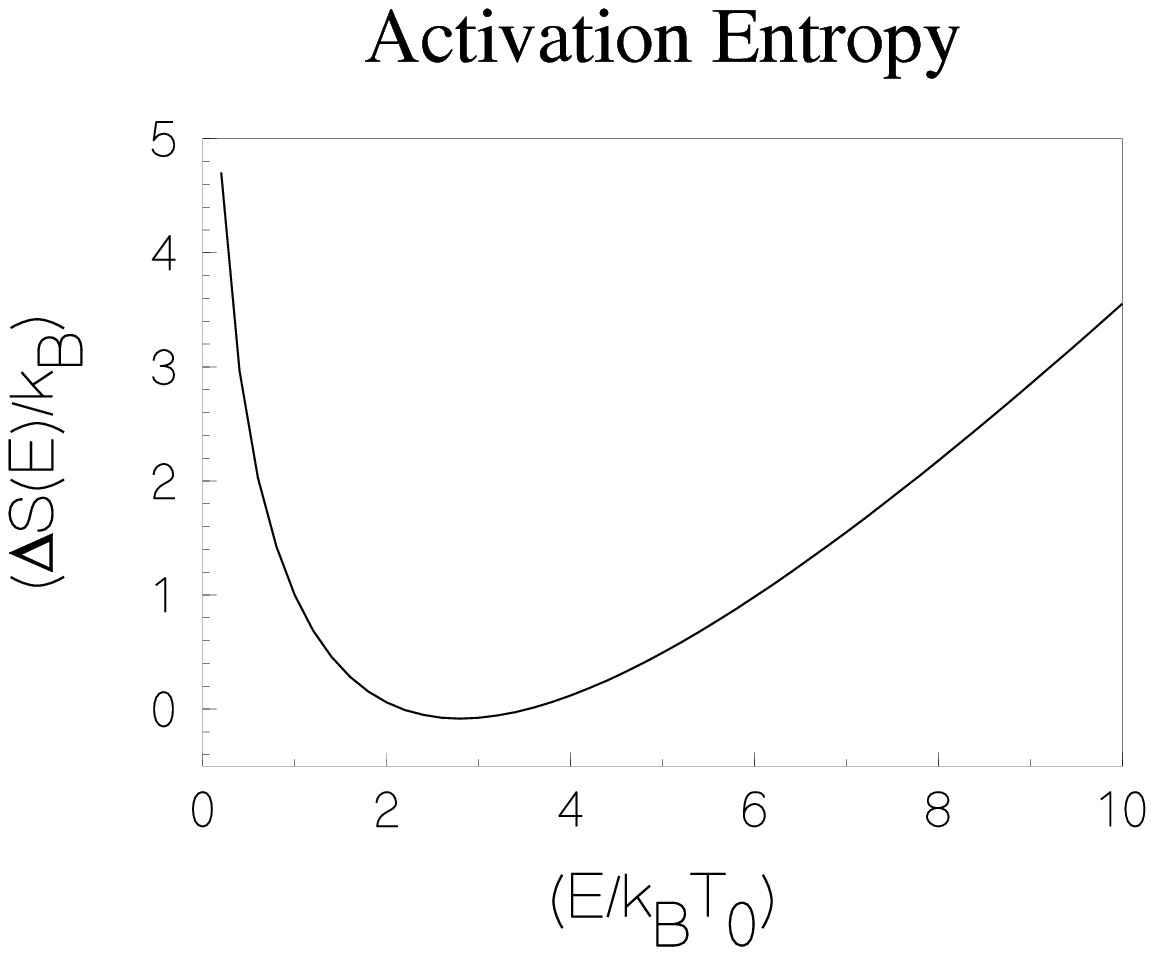}}
\caption{\it The activation entropy of mobile closed strings,  
$\Delta S(E)=S(E)-\tilde{S}- (3-\alpha) k_B\ln(\ln \mu)$, is plotted as 
a function of energy. The entropy function is convex downward indicating a metastable 
entropy activation. There is a slower rate as the entropy is 
pushed toward a minimum value.}
\label{fig2}
\end{figure}

Closed polymer chains in the form  of ``polygons'' are treated as a self 
avoiding random polygons. The number of {\em closed self avoiding polygon} polymer 
chains containing \begin{math} n \end{math} links is thought to 
obey\cite{McKenzie:1976} 
\begin{equation}
\Omega_n=\frac{\tilde{\Omega}}{n^{(3-\alpha )}}\mu^n .
\label{closed1}
\end{equation}
Wherein $\mu$ denotes the connectivity. 
The de Gennes scaling law\cite{deGennes:1979}in \begin{math} d \end{math} dimensions 
for the exponent \begin{math} \alpha \end{math} is given by 
\begin{equation}
3-\alpha =\left\{1+\frac{d}{D}\right\}
\label{closed2}
\end{equation}
wherein \begin{math} D \end{math} is the fractal dimension of the complete 
closed chain configuration. In mean field theory\cite{Flory:1953} we have 
\begin{equation}
d=3\ \ \Rightarrow \ \ D\approx \frac{5}{3}\ \  
\Rightarrow \ \ \alpha \approx \frac{1}{5}.
\label{closed3}
\end{equation}
If \begin{math} \varepsilon \end{math} denotes the activation energy per link for 
a mobile closed chain (closed string), then the energy 
\begin{equation}
E=\varepsilon n
\label{closed4}
\end{equation}
determines the entropy via Eqs.(\ref{intro3}), (\ref{closed1}) and 
(\ref{closed4}) according to 
\begin{equation}
S(E)=\tilde{S}+\frac{E}{T_0}-
(3-\alpha ) k_B \ln \left(\frac{E}{\varepsilon}\right)
\label{closed5}
\end{equation}
wherein
\begin{eqnarray}
\tilde{S} = k_B\ln \tilde{\Omega},
\nonumber \\ 
k_BT_0 = \frac{\varepsilon}{ln \mu}\ .
\label{closed6}
\end{eqnarray}
The activation entropy as a function of energy 
exhibits a minimum as shown in FIG. \ref{fig2}.
For stable entropy functions, the maximum entropy 
principle dictates upward convexity while metastable 
entropy functions exhibit downward convexity. Since 
the density of final states \begin{math} \propto e^{S/k_B} \end{math}, 
rates become slower as the minimum activation entropy is approached.

In terms of the temperature \begin{math} T \end{math},  
\begin{equation}
\frac{1}{T}=\frac{dS(E)}{dE}=\frac{1}{T_0}
-\left(\frac{(3-\alpha)k_B}{E}\right),
\label{closed7}
\end{equation}
we have at \begin{math} T_0 \end{math} the activated energy singularity
\begin{equation}
E=\left[\frac{(3-\alpha )k_BTT_0}{T-T_0}\right].
\label{closed8}
\end{equation}
Eqs.(\ref{closed3}), (\ref{closed5}) and (\ref{closed8}) imply 
\begin{eqnarray}
S&=&S_\infty+k_B\left[\frac{(3-\alpha)T_0}{T-T_0}\right] 
\nonumber \\ 
&\ &\ -(3-\alpha)k_B\ln \left[\frac{(T}{(T-T_0)}\right].
\label{closed9}
\end{eqnarray}

Using Eqs.(\ref{intro2}) , (\ref{closed3}) and (\ref{closed9}), we may now
complete the proof that the closed chain activation law has the 
VFT form given by Eq.(\ref{intro1}). Explicitly, we have
\begin{eqnarray}
\Gamma &=& \nu e^{-\Phi/k_B(T-T_0)},
\nonumber \\  
\Phi &=& (3-\alpha)k_BT_0 \approx  2.8 k_BT_0,
\nonumber \\
S_\infty &\approx & \tilde{S}+2.8 k_B
\nonumber \\
\nu &\approx & \left[\frac{(T-T_0)}{T}\right]^{2.8} 
\nu_\infty e^{S_\infty /k_B}\ .
\label{closed10}
\end{eqnarray}
In practice, the VFT activation process is often observed by 
measuring viscosity, 
\begin{equation}
\eta \approx \rho b^2 \Gamma \approx \rho b^2\nu e^{-\Phi /k_B(T-T_0)},
\label{closed11}
\end{equation}
wherein \begin{math} \rho  \end{math} is the mass density and 
\begin{math} b \end{math} is the length scale of the polymer links.
In this regard, the prediction for the activation free energy 
\begin{math} \Phi \approx 2.8 k_BT_0 \end{math}
is subject to an experimental test of the scaling critical 
index in Eq.(\ref{closed2}). 

\section{Surface Effects \label{se}}

Consider the problem of how much activation 
energy would be required to remove a given section of 
chain from the condensed matter piece of polymer. 
If the given section of chain were deep within the polymer, 
the removal would be quite difficult. For example, if one 
exerted a force on the given chain section , then it would become knotted 
with other polymer chain sections and would be rendered immobile. 
On the other hand, if the given section of chain was entirely located 
in the neighborhood of the surface boundary of the polymer, then it would 
be relatively easier to peel the chain off the surface.  

Let us consider, in more detail,  the activation energy   
to slide a section of polymer chain along a given path. 
Such an activation energy has been denoted above as  
\begin{math} \varepsilon  \end{math}
per link of the chain section. Furthermore, let 
\begin{math} z \end{math} denote the distance from a chain 
link to the boundary surface. By the above physical arguments 
we expect \begin{math} \varepsilon(z) \end{math} to sharply decrease as 
\begin{math} z\to 0 \end{math}. From Eq.(\ref{closed6}) we expect, for 
uniform connectivity (\begin{math} \ln \mu \approx {\rm const.} \end{math}),
the dynamical singularity temperature to be a decreasing function of 
\begin{math} z \end{math} varying as  
\begin{equation}
\frac{T_0(z)}{T_0}\approx \frac{\varepsilon(z)}{\varepsilon}
\ \ {\rm and}\ \ \frac{dT_0(z)}{dz}<0
\ \ {\rm as}\ \ z\to 0^+\ . 
\label{se1}
\end{equation}
In a local density theory\cite{Louis:2002}, 
\begin{math} \varepsilon (z) \end{math} 
may be parameterized by 
\begin{equation}
T_0(z)\approx T_0 \tanh^2 (z/\xi ),
\label{se2}
\end{equation}
in which the coherence length is related to the density  
\begin{equation} 
\xi \sim \rho^{-\nu (3\nu-1)}
\ \ \ {\rm wherein}\ \ \ \nu=D^{-1}\approx \frac{3}{5}\ .
\label{se3} 
\end{equation}
The \begin{math} \lim_{z\to 0^+} T_0(z)=0  \end{math} relation 
invalidates the VFT Eq.(\ref{intro1}) for the case of 
very thin polymer films.

\section{Conclusion\label{conc}}
  
A derivation has been provided for VFT activated 
transition rates in bulk polymer glasses.
Our derivation depends on the micro-canonical 
counting of the number of closed polymer chain 
configurations within the bulk glassy system. 
The configuration counting is mapped into the 
self avoiding polygon problem.  
The activation energy \begin{math} \varepsilon \end{math} per 
link determining the chain mobility also determines the 
dynamical glass transition temperature in the empirical 
VFT law. The critical indices employed are calculated as 
in Flory's theory. The chain movements also
lend strong support to  ``co-operative'' motion
inside the bulk.

It is also to be stressed that the dynamical 
glass transition temperature, \begin {math}  T_0 \end {math},
varies with the distance from the surface boundary through that a coherence 
length scale of about a few nanometers. 
This surface effect is due to the fact polymer strings localized near 
the surface boundary are more mobile than the polymer chains embedded 
in the bulk. For sufficiently thin films, the VFT activation law 
thereby becomes modified as in Eqs. (\ref{se1} - \ref{se3}) .

\end{document}